\documentclass[12pt]{article}
\usepackage{graphicx}
\usepackage{xcolor} 
\usepackage{lettrine} 
\usepackage{amsmath,amsthm,amsfonts}

\def\be{\begin{equation}}
\def\ee{\end{equation}}
\def\mbs{\boldsymbol}
\def\doe{\partial}

\def\dd{{\rm d}} % derivative d
\begin{document}

%\title{Understanding the Hamiltonian Function through Partial Legendre Transforms}
\title{Understanding the Hamiltonian Function through the Geometry of Partial Legendre Transforms}
\author{John E. Hurtado\thanks{Professor, Department of Aerospace Engineering, College of Engineering; jehurtado@tamu.edu.} \\ {\it Texas A\&M University, College Station, Texas 77843-3141}}
\date{\today}
\maketitle

\lettrine[lines=1,lhang=0,loversize=0.5]{T}{he} relationship between the Hamiltonian and Lagrangean functions in analytical mechanics is a type of duality.  The two functions, while distinct, are both descriptive functions encoding the behavior of the same dynamical system.  

Each function uses a set of generalized coordinates, $q = \{q_1, \ldots, q_n\}$, to describe the configuration of the system.  For holonomic systems we consider a minimum set of independent, unconstrained coordinates.  The Lagrangean function relates the system's dynamics in terms of the generalized coordinates and their time derivatives, whereas the Hamiltonian function relates the system's dynamics in terms of generalized coordinates and their associated momentum.  

The Lagrangean function naturally appears as one investigates the fundamental equation of classical dynamics ([1] p.~28).  The outcome of the investigation is Lagrange's classical equations,\footnote{Nielsen's form of Lagrange's equations can also naturally follow from the investigation [2].} which are $n$ second-order ordinary differential equations.  The Gibbsian, a different descriptive function, can also naturally appear in this process.  In that case, the Gibbs-Appell equations are formed which are also $n$ second-order ordinary differential equations.

From this perspective, the Lagrangean and the Gibbsian functions could each be considered {\em primary} descriptive functions and any directly associated equations of motion could be considered primary forms of the equations.  

It is not that way for the Hamiltonian.  The Hamiltonian function and Hamilton's canonical equations do not naturally or directly appear when studying the fundamental equation: they only come {\em after} Lagrange's equations fully take form.  Their appearance commonly comes through a Legendre transform of the Lagrangean function ([3] \S 2.8) or by investigating the central equation built from Lagrange's equations ([4] \S 8.2, pp.~1073--1077).  Indeed, Hamilton used a version of the central equation approach when he first introduced his ideas ([5] \S 1--3).  

Papastavridis mentions that the central equation-based derivation is ``far simpler and motivated'' when compared to the Legendre transform method ([4] p.~1076).  Nevertheless, we are motivated to revisit the Legendre transform approach and offer a more refined geometrical interpretation than what is commonly shown.  Einstein's summation convention is sometimes used in this study.  

% which form a set of $2n$ first-order ordinary differential equations in the generalized coordinates and their associated momentum.  
% From the perspective that Hamilton's function and equations come into focus after Lagrange,  

\vspace {5mm}

\lettrine[lines=1,lhang=0,loversize=0.5]{T}{he} fundamental equation of classical dynamics is $\left (  m \mbs a -  \mbs f \right ) \cdot \delta \mbs r =0 $ ([1] p.~28, [4] p.~386 ff).\footnote{For the simplicity of the current presentation, but without sacrificing  generality, we consider the motion of a single particle.}  Here, $m$ is the particle's mass, $\mbs a $ is the particle's inertial acceleration vector, $\delta \mbs r$ is the virtual displacement vector, and $\mbs f$ represents the vector sum of all impressed forces not including ideal constraint forces, which have already been discarded because they perform no virtual work.   

In the usual ways, the fundamental equation becomes the following for the case that all generalized forces stem from a potential energy function $V(t,q)$.  
\be
m {\mbs a} \cdot \frac{\doe {\mbs v}}{\doe \dot q_k} = -\frac{\doe V}{\doe q_k}, \qquad k=1, \ldots, n  \label{fund1}
\ee
From this point, a key kinematic identity together with the definition of the kinetic energy function, $T(t,q,\dot q) = \frac{1}{2} m {\mbs v} \cdot {\mbs v}$, and the introduction of the Lagrangean function, $L(t,q,\dot q) = T-V$, transmutes eq.~(\ref{fund1}) to Lagrange's classic equations.     
\be
\frac{\dd}{\dd t} \left ( \frac{\doe L}{\doe \dot q_k} \right ) - \frac{\doe L}{\doe q_k} = 0, \qquad k=1, \ldots, n \label{lagclassic}
\ee

Equation (\ref{lagclassic}) was Hamilton's starting point.  Throughout his study he considered the kinetic energy (hence, the Lagrangean) as a homogeneous quadratic form in the generalized velocities.  He lost nothing in this consideration because even the most general quadratic form can be arranged in this way ([4] p.~512).  This allowed him to use Euler's homogeneous function theorem as he explored variations, which led him to  momentum states, a dual function, and his canonical equations.

It is interesting that the familiar expression that relates the Lagrangean function and the dual function (i.e., the Hamiltonian) does not appear in his paper; it's evidently not needed in his approach.     

\vspace {5mm}

\lettrine[lines=1,lhang=0,loversize=0.5]{C}{ontinuing,} we remark that the Lagrangean is always a convex function in the generalized velocities $\dot q_k$ for holonomic systems described by a minimum set of independent, unconstrained generalized coordinates.  Convex functions of a set of independent variables enjoy the property of having a one-to-one correspondence between each independent variable and the slope of the function with respect to that variable.  In this context, that means a one-to-one relationship between each generalized velocity $\dot q_k$ and the slope (i.e., partial derivative) of the Lagrangean with respect to the corresponding generalized velocity, $s_k = \doe L / \doe \dot q_k$.  Therefore, $s_k$ is a single-valued function of $\dot q_k$ that can be inverted to give $\dot q_k$ as a single-valued function of $s_k$.  

With this understanding, one could consider exchanging each generalized velocity for its corresponding slope and rewriting the Lagrangean function as $L(t,q,\dot q(s)) = L^\ast(t,q,s)$.  Lagrange's equations would then have following appearance.  
\be
\frac{\dd}{\dd t} \left ( M_{ki} \frac{\doe L^\ast}{\doe s_i} \right ) - \frac{\doe L^\ast}{\doe q_k} = 0, \qquad k=1, \ldots, n\label{l_ast}
\ee
Here, $M_{ki}(t,q)$ is the system mass matrix, which is the Hessian matrix with respect to the generalized velocities of the Lagrangean function.  Note that eq.~(\ref{l_ast}) gives $n$ first-order ordinary differential equations in the slopes $s_k$.  The other $n$ first-order differential equations governing the motion come from $\dot q_k(t,q,s)$.  
  Equation (\ref{l_ast}) has an inadequate semblance.   
  
Exchanging each generalized velocity for its corresponding slope prompts us to consider exchanging the Lagrangean function $L(t,q,\dot q)$ for a new function that is more than the {\em function of a function} viewpoint of $L(t,q,\dot q(s)) = L^\ast(t,q,s)$.  Toward this end, the convex Lagrangean function and the one-to-one correspondence between each generalized velocity $\dot q_k$ and slope $s_k$ encourages a point-slope perspective and therefore the point-slope formula $y = m x + b$ in this context deserves attention.  

A single degree of freedom system would have the point-slope expression $L = s \dot q + b$.  In this arrangement, the slope $s$ and the $y$-intercept $b$ are taken as functions of time, the generalized coordinate, and the generalized velocity.  Inverting this expression to isolate the $y$-intercept has us treating $b$ as a function of time, the generalized coordinate, and the slope: $-b = s \dot q - L$.  The negative of this $y$-intercept is our new function.  
\be
H = s \dot q - L \label{legtra}
\ee
An illustration of this single degree of freedom case is shown on the left in fig.~1.  A representative $\dot q$ is selected for which the open circle denotes the value of $L$ and the filled circle denotes the $y$-intercept, hence value of $H$.  The value of $L$ plus the value of $H$ equals the product of the slope and point.  

Equation (\ref{legtra}) is recognized as a Legendre transform between functions $L$ and $H$ for this single degree of freedom case and this geometrical interpretation is not new ([4] p.~1076, [6]).

\begin{figure}
    \centering
    \includegraphics[width=0.75\linewidth]{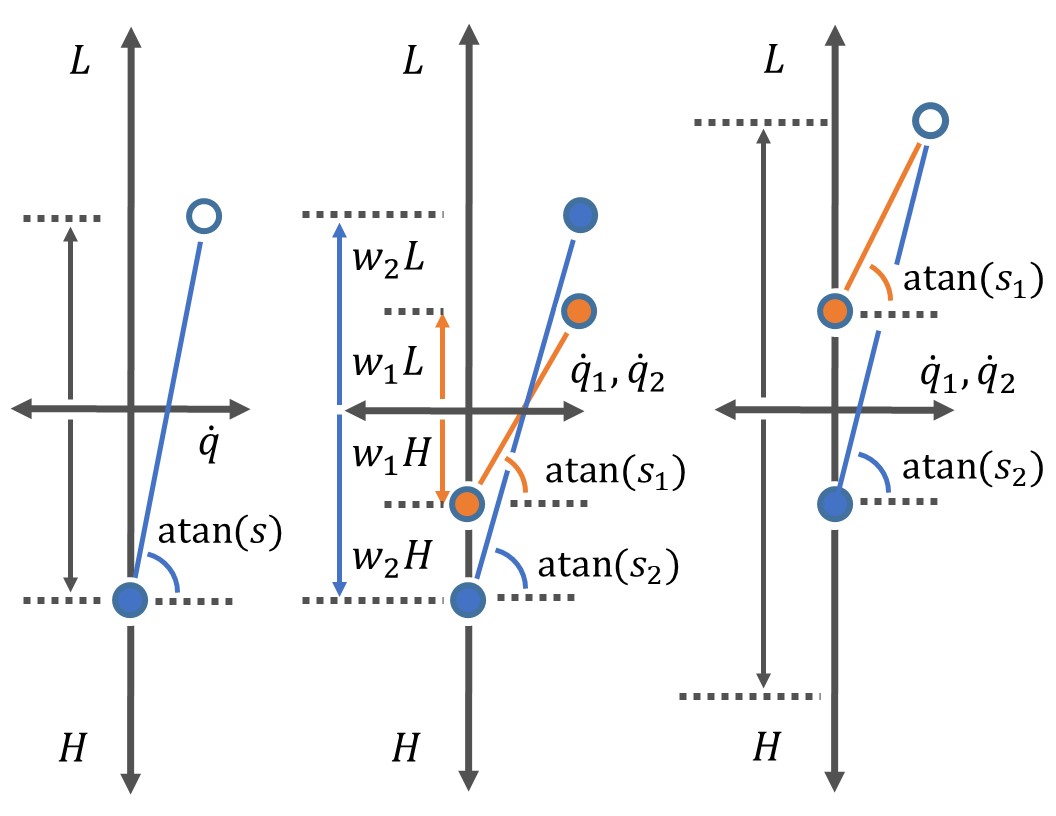}
    \caption{Left: Illustration of a Legendre transform for a single degree of freedom system; Middle: Illustration of partial Legendre transforms for a two degree of freedom system; Right: Illustration of the system Legendre transform for a two degree of freedom system.}
%    \label{fig:enter-label}
\end{figure}

To expand this interpretation to systems with $n$ generalized coordinates and velocities we propose a weighted point-slope formula for each point and slope pair.  
\be
H w_j = s_j \dot q_j - L w_j, \,\text{where $w_j \ge 0$, $j=1, \ldots, n$, and no sum on $j$} \label{indlt}
\ee
We require the weights to sum to one and propose the following form as a candidate. 
\be
w_j = \frac{\left \Vert s_j \dot q_j \right \Vert}{\sum_{i=1}^n \left \Vert s_i \dot q_i \right \Vert}, \quad \text{no sum on $j$} \label{wghts}
\ee
Each weighted point-slope expression in eq.~(\ref{indlt}) is a {\em partial} Legendre transform.  An illustration for two representative point and slope pairs is shown in the middle of fig.~1.  For each pairing, the weighted value of $L$ plus the weighted value of $H$ equals the product of the slope and point.      

The Legendre transform for the system is the summation of the partial Legendre transforms.  
\be
\sum_{j=1}^n H w_j = \sum_{j=1}^n \left ( s_j \dot q_j - L w_j \right ) \qquad \to \qquad  H = s_k \dot q_k - L \label{syslt}
\ee
An illustration for the system Legendre transform is shown on the right in fig.~1.  A representative point in $\dot q$ space is selected for which the open circle denotes the value of $L$ and each filled circle denotes a $y$-intercept.  The sum of the slope and point products equals the sum of the $L$ and $H$ function values.      

\vspace {5mm}

\lettrine[lines=1,lhang=0,loversize=0.5]{D}{iscovering} Hamilton's canonical equations from the system Legendre transform is a straightforward and well-known process, but we include it here for completeness.  Comparing the differential of eq.~(\ref{syslt}) with the differential of the function $H$ gives the results.  
\begin{equation*}
s_k \dd \dot q_k + \dot q_k \dd s_k - \frac{\doe L}{\doe q_k} \dd q_k - \frac{\doe L}{\doe \dot q_k} \dd \dot q_k - \frac{\doe L}{\doe t} \dd t
 = \dd H = \frac{\doe H}{\doe q_k} \dd q_k + \frac{\doe H}{\doe s_k} \dd s_k + \frac{\doe H}{\doe t} \dd t 
\end{equation*}
\be
\to \quad s_k =  \frac{\doe L}{\doe \dot q_k}  \quad ; \quad  \dot q_k = \frac{\doe H}{\doe s_k} \quad ; \quad - \frac{\doe L}{\doe q_k} = \frac{\doe H}{\doe q_k} \quad ; \quad  - \frac{\doe L}{\doe t} = \frac{\doe H}{\doe t} \label{coll}
\ee
These expressions can be used in Lagrange's classic equations eq.~(\ref{lagclassic}) as needed to give $\dot s_k = - {\doe H}/{\doe q_k}$. This and $\dot q_k = {\doe H}/{\doe s_k}$ from eq.~(\ref{coll}) are Hamilton's canonical equations of motion.  

Finally, the slopes $s_k = {\doe L}/{\doe \dot q_k}$ are more conventionally known as momenta; we used the term ``slope'' simply because of the geometric focus of this study.  

% $p_k = \doe L / \doe \dot q_k$   %papa says that Tabor p79-80 has an alternative geometrical interpretation.  

\section*{References}
\begin{enumerate}

\item [{[1]}] Pars, L.A., {\em A Treatise on Analytical Dynamics}, Ox Bow Press, Woodbridge, CT, 1981.  

\item [{[2]}] Hurtado, J.E., ``New Time-Integral Variational Principle,'' submitted to {\em The Journal of Astronautical Sciences}, 16 OCT 2023. 

\item [{[3]}] McCauley, J.L., {\em Classical Mechanics: transformations, flows, integrable, and chaotic dynamics}, Cambridge University Press, New York, NY, 1997.  

\item [{[4]}] Papastavridis, J.G., {\em Analytical Mechanics}, Oxford Univ.~Press, New York, NY, 2002.  %See papa p400 for momentum in AM comment.  

\item [{[5]}] Hamilton, W.R., ``Second Essay on a General Method in Dynamics,'' {\em Philosophical Transactions of the Royal Society}, Vol.~125, pp.~95-144.

\item [{[6]}] Zia, R.K., Redish, E.F., and McKay, S.R., ``Making Sense of the Legendre Transform,'' {\em American Journal of Physics}, Vol.~77, Iss.~7, 2009.  

\end{enumerate}
\end{document}